\documentclass[prb,twocolumn,floatfix,showpacs,amssymb]{revtex4}
\usepackage{graphicx}
\usepackage{dcolumn}
\usepackage{bm}
\begin{document}

\title{Crossed conductance in FSF double junctions:\\
role of out-of-equilibrium populations}
\author{R. M\'elin}
\affiliation{
Centre de Recherches sur les Tr\`es Basses
Temp\'eratures (CRTBT\footnote{U.P.R. 5001 du CNRS, Laboratoire
conventionn\'e avec l'Universit\'e Joseph Fourier}),\\ CNRS, BP 166,
38042 Grenoble Cedex 9, France}
\begin{abstract}
We discuss a model of Ferromagnet / Superconductor / Ferromagnet (FSF)
double junction
in which the quasiparticles are not in equilibrium
with the condensate in a region of the superconductor containing
the two FS contacts. The role of geometry is discussed, as
well as the role of a small residual density of states within
the superconducting gap, that allows a sequential tunneling crossed
current.
With elastic quasiparticle transport and the geometry with lateral contacts,
the crossed conductances in the sequential tunneling channel are almost
equal in the normal and superconducting phases,
if the distance between the FS interfaces is sufficiently small. 
The sequential tunneling and spatially separated processes 
(the so-called crossed Andreev reflection and elastic cotunneling
processes) lead to
different signs of the crossed current in the antiparallel alignment
for tunnel interfaces.
\end{abstract}


\maketitle

\section{Introduction}

Transport properties of multiterminal hybrid structures
involving a superconductor (S), connected to several ferromagnets (F)
or normal metals (N)~\cite{Lambert1,Jedema}
has focused a considerable
interest recently.
A superconductor is a condensate of Cooper pairs with an energy
gap $\Delta$ to
the first quasiparticle excitations~\cite{Tinkham}. 
In FSF double
tunnel junctions,
interesting phenomena come into play
when out-of-equilibrium spin populations can be generated in the
superconductor~\cite{Takahashi,Zheng,Bozovic,Yamashita,Brataas}.
For instance
out-of-equilibrium effects have a strong influence on
the value of the self-consistent superconducting
gap of a FSF
trilayer~\cite{Takahashi,Zheng,Bozovic,Yamashita,Brataas},
that can be controlled by an applied voltage.

\begin{figure}
\includegraphics [width=1. \linewidth]{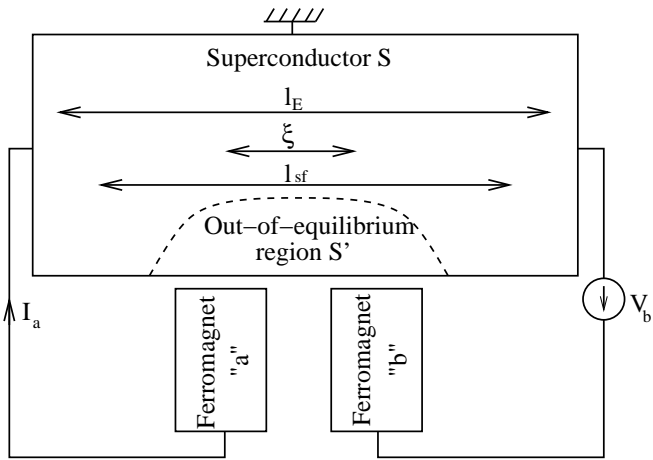}
\caption{Schematic representation of a FSF double junction with
lateral contacts, with an out-of-equilibrium region S'.
A voltage $V_a=0$ is applied on the ferromagnet ``a'', a voltage
$V_b=V$ is applied on the ferromagnet ``b'', and voltage
$V_{S}=0$ is applied on the superconductor
S. We note $N_S$ the number of channels between S and S',
and $N_F$ the number of channels in each contact between the
superconductor and the ferromagnets. 
Different length scales are indicated on the figure:
the spin-flip length $l_{sf}$ 
the superconducting coherence length $\xi$ and the energy
relaxation length $l_E$. Other situations, such
as the one with $l_E$
being the smallest length scale, are considered in the
text.
The situation on the figure corresponds
to a distance $R$ between the contacts such that
$\xi<R<l_{sf}$, in which case there is a finite crossed
current but no magnetoresistive effect.
The size of the out-of-equilibrium region is discussed in
Section~\ref{sec:region}.
\label{fig:schema}
}
\end{figure}

Cooper pairs,
being bound states of two electrons with opposite spins,
have a spatial extent $\xi$ given by the BCS coherence length.
The limit of
equilibrium transport in FSF double junctions where
the distance $R$ between the contacts
becomes smaller than
$\xi$~\cite{Lambert2,Deutscher,Falci,Melin-JPCM,Melin-Peysson,Chte,AB,Feinberg-des,Buttiker,Pistol,MF-PRB,Beckmann,Russo},
has been intensively investigated recently in
connection with the determination of the so-called ``crossed conductance''.
A voltage-biased crossed conductance experiment
similar to the ones by Beckmann {\it et al.}~\cite{Beckmann}
in the geometry with lateral contacts on Fig.~\ref{fig:schema}
consists in measuring the current $I_a$ in electrode ``a'' in response to
a voltage $V_b=V$ on electrode ''b'' while a
voltage $V_S$ is applied on the superconductor.
The trilayer geometry with extended interfaces and with tunnel
contacts was used in a
recent experiment by Russo {\it et al.}\cite{Russo}
The crossed conductance is
defined by~\cite{Falci}
\begin{equation}
\label{eq:Gab}
{\cal G}_{a,b}(V_a,V_b) = \frac{\partial I_a}{\partial V_b} 
(V_a,V_b)
,
\end{equation}
and we focus here on the case $V_a=V_S$.
Since one voltage can be chosen as a reference we use
$V_a=V_S=0$.
The proposed interpretation of the experiment by Beckmann {\it et
al.}\cite{Beckmann} and Russo {\it et al.}\cite{Russo} involves
crossed Andreev reflection and elastic
cotunneling~\cite{Lambert2,Deutscher,Falci,Melin-JPCM,Melin-Peysson,Chte,AB,Feinberg-des,Buttiker,Pistol,MF-PRB},
corresponding to transmission over two spatially separated contacts,
in the electron-hole and electron-electron channels respectively,
without out-of-equilibrium spin populations.

Beckmann {\it et al.}~\cite{Beckmann} already explained that
the magnetoresitive effects at temperatures 
comparable to the transition temperature of the superconductor
could be explained by charge and spin imbalance~\cite{Clarke}
due to out-of-equilibrium spin populations in the superconductor.
This suggests two different explanations of the magnetoresistive
effect: one explanation close to the superconducting transition
temperature based on out-of-equilibrium
populations, and one explanation in the superconducting state, based
on spatially separated processes at equilibrium (the so-called
crossed Andreev reflection and elastic cotunneling processes). 
The goal of our article is to start from a description of
magnetoresistive effects in the normal state, based on
out-of-equilibrium populations, and include
superconducting correlations. We show that 
out-of-equilibrium populations can even play a role in
subgap transport in the geometry with lateral contacts.

More specifically we
assume the existence of a region S' in the superconductor
where the quasiparticles are
not in equilibrium with the condensate (see Fig.~\ref{fig:schema}).
For highly transparent FS interfaces,
the geometry is characterized by a parameter $r=N_F/N_S$,
where $N_S$ and $N_F$ are the number of channels involved 
in the SS' and SF contacts (see Fig.~\ref{fig:schema}).
The superconductor is characterized by a residual density of
states within the superconducting gap $\rho_N \eta / \Delta$,
where $\rho_N$ is the normal density of states, and $\eta/\Delta$
is a phenomenological dimensionless parameter estimated
around $10^{-2}$ in Ref.~\onlinecite{Cuevas}, and around
$10^{-4}$ in Ref.~\onlinecite{Pekola}. This small density of
states within the superconducting gap allows a new conduction
channel (sequential tunneling) that is complementary to the
channels of spatially separated processes.
We show that the voltage
dependence of the sequential tunneling
crossed conductance depends qualitatively
on whether $\eta/\Delta \ll r \ll 1$ or
$r \ll \eta/\Delta \ll 1$.
Given the orders of magnitude
of $\eta/\Delta$\cite{Cuevas,Pekola} and the geometry of the experiment by
Beckmann {\it et al.}\cite{Beckmann}, we conclude that
the relevant regime is $r \ll \eta/\Delta$. In this case,
the sequential tunneling linear crossed conductance 
in the superconducting state is almost equal to the linear
crossed conductance in the normal state, which is 
apparently compatible
with experiments\cite{Beckmann}.

The article is organized as follows. Preliminaries are given in
section~\ref{sec:prelim}. The cases of normal and
superconducting states with tunnel junctions are presented
in section~\ref{sec:norm}. Numerical simulations
with weak and strong energy relaxations
and arbitrary interface transparencies
are presented in section~\ref{sec:num}.
Concluding remarks are given in
section~\ref{sec:conclu}. Some details
are given in the Appendix.

\section{Preliminaries}
\label{sec:prelim}

\subsection{Out-of-equilibrium region}
\label{sec:region}
We suppose the existence of
an out-of-equilibrium region S' in the superconductor
that contains the 
contacts with the two ferromagnets ``a'' and ``b''
(see Fig.~\ref{fig:schema}). The out-of-equilibrium
quasiparticle populations in the superconductor are expected
to decay over a length scale $l_0$. The length $l_0$ can have
an origin intrinsic to the superconductor, in which case it can
be identified to 
the smallest between i) the spin-flip
length $l_{sf}$, ii) the inelastic scattering length $l_{e-e}$,
iii) the recombination
length $l_R$ after which quasiparticles recombine to form
Cooper pairs\cite{Clarke}.
The length scale $l_0$ can also be due to the inverse
proximity effect with the ferromagnets. The density of states induced
in the superconductor due to the inverse
proximity effect at a FS tunnel interface is proportional
to $1/R^2 \exp{(-R/\xi)}$ for a ballistic system, and to
$1/R \exp{(-R/\xi)}$ for a diffusive system\cite{Feinberg-des}, where
$R$ is the distance to the FS contact and $\xi$ the coherence length.
There can thus exist a small but finite density of states decaying
algebraically up to distances
comparable to the superconducting coherence length. 
Treating the full spatial
dependence of the quasiparticle populations and
density of state goes beyond the scope of our article.
Instead we replace
them by a step-function variation so that
there exists a region S' with uniform non vanishingly small
quasiparticle potentials, connected to the remaining of the
superconductor S in which the quasiparticles are in equilibrium
with the condensate.

The number of channels $N_S$ connecting the out-of-equilibrium
region S'
to the remaining of the superconductor S is supposed to be large
enough so that the phase and the chemical potential of the
condensate are identical in S and S', but the quasiparticle
populations can be different in S and S'.
Moreover the SS' contacts are highly
transparent.
Each of the FS interfaces is supposed to
contain $N_F$ channels.
The SF contacts can be highly transparent, like in the experiment
by Beckmann {\it et al.}~\cite{Beckmann}, or have a small
transparency.

The relevant parameter characterizing the geometry of the
FSF double junction is
$r=N_F / N_S$. The trilayer geometry used by
Russo {\it et al.}\cite{Russo} corresponds to
$r\agt 1$ while the geometry with lateral contacts
used by Beckmann {\it et al.}\cite{Beckmann}
corresponds to $r \ll 1$.

\subsection{Hamiltonians}

The superconductor is described by the BCS Hamiltonian~\cite{Tinkham}
\begin{eqnarray}
\label{eq:H-BCS}
{\cal H}_{\rm BCS} &=& \sum_{\langle \alpha , \beta
\rangle , \sigma} - t \left(
c_{\alpha,\sigma}^+ c_{\beta,\sigma}
+c_{\beta,\sigma}^+ c_{\alpha,\sigma} \right)\\
&+& \Delta \sum_\alpha \left(
c_{\alpha,\uparrow}^+ c_{\alpha,\downarrow}^+
+ c_{\alpha,\downarrow}
c_{\alpha,\uparrow} \right)
.
\nonumber
\end{eqnarray}
The ferromagnets are described by
the Stoner model
with an exchange field $h_{\rm ex}$:
\begin{eqnarray}
\label{eq:H-Stoner}
{\cal H}_{\rm Stoner} &=& \sum_{\langle \alpha,\beta
\rangle,\sigma} -t \left( c_{\alpha,\sigma}^+
c_{\beta,\sigma} + c_{\beta,\sigma}^+ 
c_{\alpha,\sigma} \right)\\
&-&h_{\rm ex} \sum_\alpha
\left( c_{\alpha,\uparrow}^+ c_{\alpha,\uparrow}
-c_{\alpha,\downarrow}^+ c_{\alpha,\downarrow} \right)
\nonumber
,
\end{eqnarray}
where we supposed for simplicity that the bulk hopping
amplitudes $t$ are identical in the superconductor
and ferromagnets.
We introduced in Eqs.~(\ref{eq:H-BCS}) and (\ref{eq:H-Stoner})
a cubic lattice with discrete ``sites'' labeled by $\alpha$
and $\beta$. The symbol $\langle \alpha,\beta \rangle$ 
denotes neighboring sites on this cubic lattice,
and $\sigma=\uparrow,\downarrow$ is the component of the
spin along the $z$ axis.
The contact
between the superconductor and ferromagnet ``a'' is
described by a tunnel Hamiltonian with a hopping amplitude
$t_a$:
\begin{equation}
\label{eq:W}
{\cal W}_a = t_a \sum_k \left(
c_{\alpha_k}^+ c_{a_k} + c_{a_k}^+ c_{\alpha_k} \right)
,
\end{equation}
where the summation runs over all sites at the interface.
A site $\alpha_k$ on the superconducting side of the interface
corresponds to a site $a_k$ in the ferromagnet ``a''.
An expression
similar to Eq.~(\ref{eq:W}) is used at the interface with the
ferromagnet ``b''.
The interface transparencies are parameterized 
by $\tau_{a,b}=\pi t_{a,b} \rho_0$, where $\rho_0$ is
the normal density of state $\rho_N$ if the superconductor, taken
equal to the density of state $\rho_F$ of the ferromagnet without
spin polarization.
Highly transparent interfaces
correspond to $\tau_{a,b}=1$. The transparency of the SS'
contact is such that $\tau_S=1$.

\subsection{Green's function method}
\subsubsection{Transport formula}

The currents are obtained by evaluating the advanced ($\hat{G}^A$),
retarded ($\hat{G}^R$) and 
Keldysh ($\hat{G}^{+,-}$) Green's functions~\cite{Caroli,Cuevas}.
The Green's functions in the superconductor and ferromagnets
correspond to the continuum limit since we are interested only
in energies close to the Fermi energy. Nevertheless we introduce 
a discrete lattice to define the tunneling term (\ref{eq:W})
of the Hamiltonian.

The advanced and retarded
Nambu Green's functions are obtained by inverting the Dyson equation,
that, in a compact notation, takes the form
$\hat{G}=\hat{g}+ \hat{g} \otimes \hat{\Sigma} \otimes \hat{G}$,
where $\hat{g}$ is the Nambu Green's function of an isolated
electrode, $\hat{\Sigma}$ is the self-energy corresponding to the
tunnel Hamiltonian (\ref{eq:W}) and $\hat{G}$ is the Green's function
of the connected structure. The symbol $\otimes$ denotes a summation
over the lattice sites involved in the self-energy.
Transport properties are obtained by evaluating the Keldysh
Green's function $\hat{G}^{+,-}$ given by
the Dyson-Keldysh equation
\begin{equation}
\hat{G}^{+,-}
= [\hat{I}+\hat{G}^R \otimes \hat{\Sigma}]
\otimes \hat{g}^{+,-} \otimes
[\hat{I}+ \hat{\Sigma} \otimes \hat{G}^A ]
.
\end{equation}
The current in the sector $S_z=1/2$ (corresponding to a spin-up electron
or a spin-down hole) flowing between the two lattice
sites $a$ and $\alpha$, is given by
\begin{equation}
I_{a,\alpha}=\frac{e}{2 h} \int d \omega
\mbox{Tr}
\left\{ \left[\hat{t}_{a,\alpha} \hat{G}^{+,-}_{\alpha,a}(\omega)
- \hat{t}_{\alpha,a} \hat{G}^{+,-}_{a,\alpha}(\omega) \right]
\hat{\sigma}^z \right\},
\label{eq:I-a-al}
\end{equation}
where $\hat{\sigma}^z$ is one of the Pauli matrices, and
the trace is a summation over the ``11'' and ``22'' Nambu
components,
corresponding to spin-up electrons and spin-down holes
respectively.

The Andreev current between the ferromagnet ``a'' and S'
is vanishingly small since the voltage $e V_a$ is equal to the
pair chemical potential in S'. The quasiparticle current
is finite because of the non equilibrium 
populations in S'. The current due to 
the spatially separated processes
between S and the ferromagnet ``a'' is vanishingly small
since S and the ferromagnet ``a'' are in equilibrium,
with the same chemical potentials.

\subsubsection{Local Green's functions}
The local 
advanced Green's function of a
ferromagnet with a polarization  $P$ is given by
\begin{equation}
\label{eq:g-ferro}
\hat{g}=i \pi \rho_F \left[\begin{array}{cc} 1+P & 0 \\
0 & 1-P \end{array} \right]
.
\end{equation}
We discard the energy dependence of the ferromagnet Green's functions
since we consider energies much smaller than the exchange field.

The local advanced Nambu Green's function
$\hat{g}_S$ of an isolated superconductor takes the form
\begin{equation}
\label{eq:gS-loc}
\hat{g}_S(\omega)=\frac{\pi \rho_N}{\sqrt{\Delta^2-
(\omega-i\eta)^2}}
\left[ \begin{array}{cc} -\omega+i\eta & \Delta \\
\Delta & -\omega+i\eta \end{array} \right]
,
\end{equation}
where $\omega$ is the energy with respect to the equilibrium
chemical potential,
$\rho_N$ is the normal state density of states, and
$\eta$ is a small phenomenological
energy relaxation parameter~\cite{Cuevas,Kaplan,Pekola},
the origin of which can be intrinsic to the superconductor
(inelastic electron-electron interaction that dominate over
inelastic phonon processes at low temperature~\cite{Kaplan})
or extrinsic (the inverse proximity effect). We have
$\eta=\eta_{\rm ext}+\eta_{\rm int}$, with
$\eta_{\rm int}$ the intrinsic value of $\eta$, and
$\eta_{\rm ext}$ the extrinsic value. 
The parameter
$\eta$ estimated to $\eta/\Delta=10^{-4}$
was introduced recently~\cite{Pekola}
as a limitation to the cooling power of microfridges based on NS
junctions. The estimate $\eta/\Delta=10^{-2}$ can be found in
Ref.~\onlinecite{Cuevas}. We will use
$\eta/\Delta=10^{-2}$ and $\eta/\Delta=10^{-3}$ in what follows.
The final results do not depend crucially on the precise
of the absolute
value of $\eta/\Delta$, but rather on how $\eta/\Delta$
compares to $r$.
The density of states at zero energy is
\begin{equation}
\label{eq:rhoS}
\rho_S(\omega=0)=\frac{1}{\pi}
\mbox{Im} [ g_S^{1,1}(\omega=0) ] \simeq
\rho_N \eta / \Delta
,
\end{equation}
reduced by a factor $\eta/\Delta$
compared to the normal state density of states.

\section{Crossed current in the normal and superconducting states}
\label{sec:norm}
We start by describing magnetoresistive
effects in the situations of linear response where either
a voltage $e V_b \ll \Delta$ is applied on the ferromagnet ``b'', or
S and S' are in the normal state.
We note $\tau_E$ the energy relaxation time,
$\tau_d$ the transport dwell time (the average time spent by a
quasiparticle in S') and $\tau_{sf}$ the spin-flip time.
We examine the two cases
$\tau_E \ll \tau_d \ll \tau_{sf}$
and $\tau_d \ll \tau_E \ll \tau_{sf}$, as well as the
case of strong spin-flip $\tau_{sf}\ll \tau_d, \tau_E$.
Even though not directly relevant to the experiments
by Beckmann  {\it et al.}\cite{Beckmann}, we examine also
the hypothesis $\tau_E \ll \tau_d$ that was used recently
in the study of spin imbalance in the FSF
trilayer\cite{Takahashi,Zheng,Bozovic,Yamashita,Brataas}.
The transport dwell time
$\tau_d$
is larger for small interface transparencies in the
trilayer geometry, so that
$\tau_d$ can possibly exceed $\tau_E$ in this situation.

In the normal state
the spin-$\sigma$ current from the ferromagnet
``a'' to S', the ferromagnet ``b'' to S', and from
S to S' are given by
\begin{eqnarray}
\nonumber
I_{a\rightarrow S'}^{(\sigma)} &=& \frac{e}{h} N_F
T_a^{(\sigma)} \int d \omega
\left[ f_{S'}^{(\sigma)}(\omega) - n_F(\omega-e V_a) \right] \\
\nonumber
I_{b\rightarrow S'}^{(\sigma)} &=& \frac{e}{h} N_F
T_b^{(\sigma)} \int d \omega
\left[ f_{S'}^{(\sigma)}(\omega) - n_F(\omega-e V_b)  \right] \\
\label{eq:ISSprime}
I_{S\rightarrow S'} &=& \frac{e}{h} N_S
T_S \int d \omega
\left[  f_{S'}^{(\sigma)}(\omega) -  n_F(\omega)\right] 
,
\end{eqnarray}
where $f_{S'}^{(\sigma)}(\omega)$ is the 
spin-$\sigma$ distribution function in S',
$n_F(\omega)$ is the Fermi distribution function
at zero temperature, and
where the transmission coefficients $T_a^{(\sigma)}$, $T_b^{(\sigma)}$
and
$T_S$ are supposed to energy-independent (the full energy dependence
will be treated by numerical simulations in section~\ref{sec:num}).

\subsection{Weak energy relaxation ($\tau_d \ll \tau_E \ll \tau_{sf}$)}
\label{sec:normal-elas}
Assuming spin conserving
elastic incoherent transport (corresponding to
$\tau_d \ll \tau_E \ll \tau_{sf}$), we impose
current conservation for each energy to obtain
\begin{eqnarray}
f_S^{(\sigma)}(\omega) &=& 
\frac{T_a^{(\sigma)} N_F}{\cal D^{(\sigma)}} n_F(\omega-e V_a) \\ 
\nonumber
&+&\frac{T_b^{(\sigma)} N_F}{\cal D^{(\sigma)}} n_F(\omega-e V_b) 
+ \frac{T_S N_S}{\cal D^{(\sigma)}} n_F(\omega)
,
\end{eqnarray}
with ${\cal D}^{(\sigma)}=
(T_a^{(\sigma)}+T_b^{(\sigma)})N_F+T_S N_S$.

Considering a geometry with lateral contacts used by 
Beckmann {\it et al.}\cite{Beckmann}, and the magnitude
of $\eta / \Delta$\cite{Cuevas,Pekola},
$N_S$ is so huge that $T_S N_S \gg T N_F$, both in the normal 
and superconducting states. This means
$r=N_F/N_S \ll r_*$, where the cross-over value of $r$
is given by
\begin{equation}
r_*=\frac{t_S^2}{t_F^2} \frac{\eta_{\rm int}+\eta_{\rm ext}}{\Delta}
.
\end{equation}
Taking
an estimate of $\eta_{\rm ext}/\Delta$ in the tunnel limit for the
FS contact and in the diffusive limit for the
superconductor\cite{Feinberg-des}, we obtain
\begin{equation}
\frac{\eta_{\rm ext}}{\Delta}
\simeq N_F\frac{t_F^2 \rho_F \rho_N}
{k_F^2 l_e R} \exp{(-2 R / \xi)}
,
\end{equation}
from what we deduce
\begin{equation}
\label{eq:r_r*}
\frac{r}{r_*} \simeq \pi^2
k_F^2 l_e \frac{R}{N_S} \exp{(2 R / \xi)}
,
\end{equation}
where $\xi$ is the superconducting coherence length, and
where we supposed $\eta_{\rm int}\ll \eta_{\rm ext}$
We obtain a cross-over from $r \ll r_*$ for
$R\ll N_S$ (corresponding to a point in the superconductor
close to the contacts)
to $r \simeq r_*$ at the cross-over $R \simeq N_S$. The exponential
increase for $R > \xi$ is cut-off by the
intrinsic value $\eta_{\rm int}/\Delta$
of $\eta/\Delta$, not taken into account
in Eq.~(\ref{eq:r_r*}).

The total currents flowing from $a$ to S'
in the parallel (P) and antiparallel (AP) alignments
are given by
\begin{eqnarray}
\label{eq:current-P}
I_{S'\rightarrow a}^{tot,P} &=& -\frac{e^2}{h} N_F^2  
\left[ \frac{T^2}{2 T N_F+T_S N_S} \right.\\
&+& \left. \frac{t^2}{2 t N_F+T_S N_S} \right]
V_b \nonumber \\
\label{eq:current-AP}
I_{S'\rightarrow a}^{tot,AP} &=& -\frac{e^2}{h} N_F^2
\frac{2 T t}{(T+t)N_F+T_S N_S} V_b
,
\end{eqnarray}
where $T$ and $t$ denote 
the transmission coefficient of majority 
and minority spins. 
We supposed $r \ll r_*$ in the derivation of
Eqs.~(\ref{eq:current-P}) and (\ref{eq:current-AP}).
The crossed current is
{\it negative}, and larger in absolute value in the 
parallel alignment. 

The transmission coefficients
$T$ and $t$ are both proportional to $\eta/\Delta$, while 
$T_S$ is proportional to
$(\eta/\Delta)(\eta_{\rm int}/\Delta)$. 
Assuming that $\eta_{\rm int}$ and $\eta_{\rm ext}$ have roughly
the same order of magnitude, we conclude that
the factors of order
$\eta/\Delta$ simplify between the numerator and denominator of
Eqs.~(\ref{eq:current-P}) and (\ref{eq:current-AP}) in the
limit $T_S N_S \gg T N_F$. As a consequence, the crossed current
takes approximately the same value in the situations where the
superconductor is in the normal and superconducting states,
which is compatible with
the experiments in Ref.~\onlinecite{Beckmann}.

\begin{figure*}
\includegraphics [width=1. \linewidth]{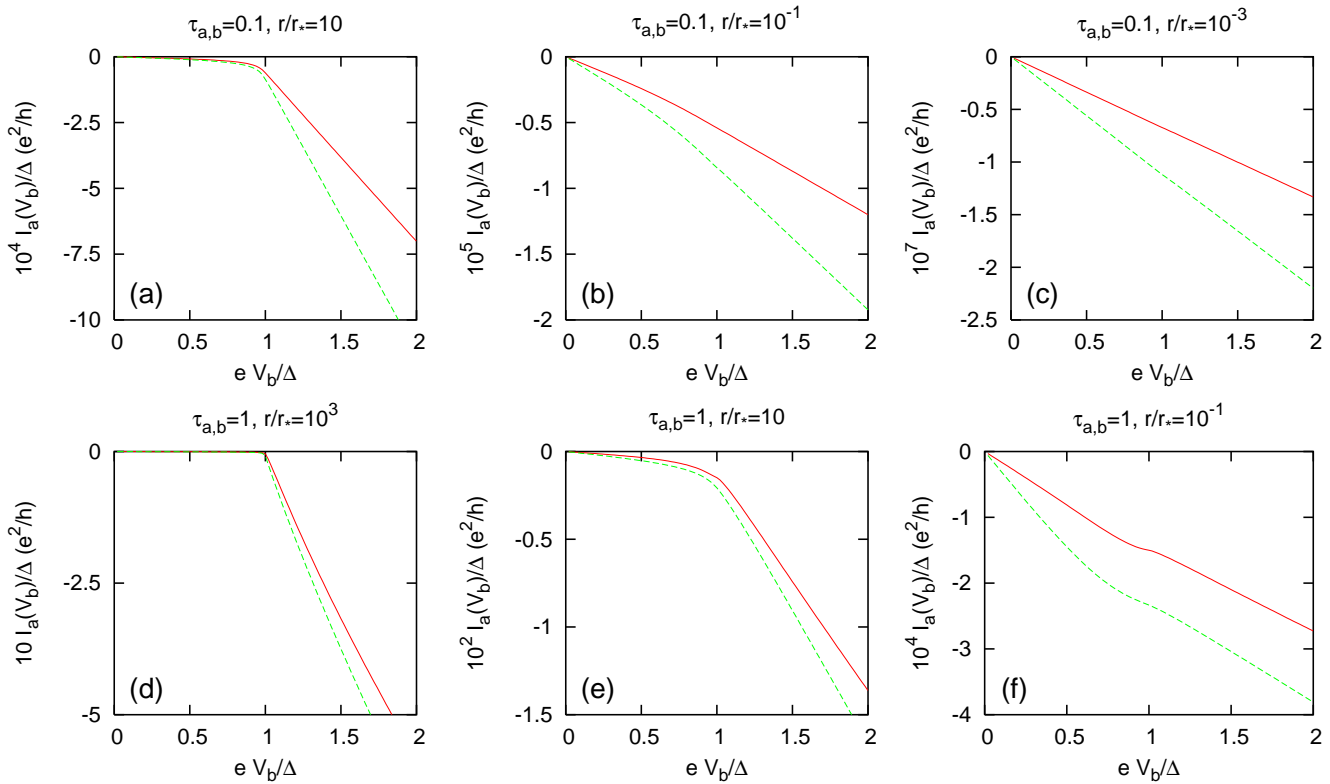}
\caption{Variation of the normalized crossed current
$I_a(V_b)/\Delta$
as a function of the normalized voltage $e V_b/\Delta$ applied
on electrode ``b'', for $P=0.5$. The solid line (red) corresponds
to the antiparallel alignment and the dashed line (green) to the parallel
alignment. The absolute value of the crossed current is larger
in the parallel alignment.
Panels (a), (b) and (c) correspond to $\tau_{a,b}=0.1$,
and $r/r_*=10^3$ (a), $r/r_*=10$ (b), and $r/r_*=10^{-1}$ (c).
Panels (d), (e) and (f) correspond to $\tau_{a,b}=1$,
and $r/r_*=10^3$ (d), $r/r_*=10$ (e), and $r/r_*=10^{-1}$ (f).
We use $\eta/\Delta=10^{-3}$.
}
\label{fig:elas-P}
\end{figure*}

\subsection{Strong energy relaxation ($\tau_E \ll \tau_d \ll \tau_{sf}$)}
\label{sec:Vqp-nor}
\label{sec:norm-V}
If we suppose strong energy relaxation ($\tau_E \ll \tau_d \ll \tau_{sf}$),
the quasiparticle
distribution functions in S' are given
by the Fermi distribution with quasiparticle potentials
$V_{qp}^\uparrow$ and $V_{qp}^\downarrow$ for spin-up and spin-down
electrons. The total current can then be calculated as an integral over energy
of the transmission coefficient, and Kirchoff laws can be imposed
on the integrated current.
The spin-$\sigma$ quasiparticle potential $V_{qp}^{\sigma}$ is
given by
\begin{equation}
V_{qp}^\sigma = \frac{T_b^\sigma N_F}
{(T_a^\sigma + T_b^\sigma) N_F+ T_S N_S} V_b
,
\end{equation}
and the current flowing from S' to the ferromagnet ``a'' in
the parallel and antiparallel alignments are given by
the same expressions as
in section~\ref{sec:normal-elas}.

\subsection{Strong spin flip ($\tau_{sf}\ll \tau_E ,\tau_d$)}
Increasing the distance between the ferromagnetic electrodes
tends to increase the transport
dwell time, that can become larger
than the spin flip length.  
The magnetoresistive effect in the
crossed current decays exponentially
as a function of the distance between the contacts,
on a length scale set by the spin-flip length. The spin-flip
length is reduced by superconducting correlations
\cite{Belzig}, so that the crossed current
in the superconducting state is reduced, as compared to the normal
case. This effect is compatible with experiments \cite{Beckmann}.

In the case of strong spin-flip with energy conservation
($\tau_{sf}\ll \tau_d \ll \tau_E$), and without
energy conservation ($\tau_{sf}\ll \tau_E \ll \tau_d$),
the crossed current takes the form
\begin{equation}
I_{S'\rightarrow a}=-2 \frac{e^2}{h}
N_F^2
\frac{T_a T_b}{(T_a+T_b)N_F+T_S N_S}
,
\end{equation}
with $T_a=(T_a^\uparrow
+T_a^\downarrow)/2$ and
$T_b=(T_b^\uparrow+T_b^\downarrow)/2$.
The crossed current due to out-of-equilibrium populations
is negative, but remains finite,
even though there is no magnetoresistive effect.
The crossed current due to the spatially separated
process tends to zero in the limit where the
distance between the contacts is large compared to the
superconducting coherence length.
This may be used in experiments to determine whether the large
distance behavior is due to out-of-equilibrium spin populations,
or to the spatially separated processes.

\begin{figure*}
\includegraphics [width=1.\linewidth]{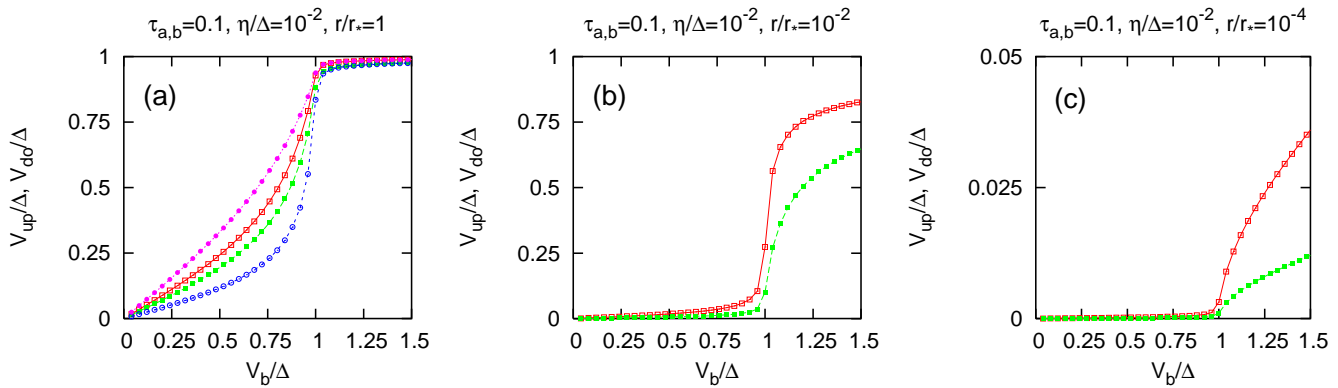}
\caption{(a) Variation of the normalized
quasiparticle potentials $V_{qp}^\uparrow(V_b)/\Delta$
and $V_{qp}^\downarrow(V_b)/\Delta$ as a function of $V_b/\Delta$, for
$\eta/\Delta=10^{-2}$, $\tau_{a,b}=0.1$. The three panels
correspond to $r/r_*=1$ (a), $r/r_*=10^{-2}$ (b), and $r/r_*=10^{-4}$ (c).
The different curves correspond to $V_{qp}^\uparrow(V_b)$
in the parallel (P) alignment (red, $\Box$), and 
to $V_{qp}^\downarrow(V_b)$
in the P alignment (green, $\blacksquare$).
$V_{qp}^\uparrow(V_b)$ is larger than $V_{qp}^\downarrow(V_b)$.
On panel (a) we have shown
to $V_{qp}^\uparrow(V_b)$
in the antiparallel (AP) alignment (blue, $\circ$),
and to $V_{qp}^\downarrow(V_b)$
in the antiparallel (AP) alignment (violet, $\bullet$).
We shown on the P case on panels (b) and (c) since
$V_{qp}^{\uparrow,P}(V_b)/\Delta
\simeq V_{qp}^{\downarrow,AP}(V_b)/\Delta$
and $V_{qp}^{\uparrow,AP}(V_b)/\Delta
\simeq V_{qp}^{\downarrow,P}(V_b)/\Delta$ in these cases.
}
\label{fig:Vqp}
\end{figure*}

\begin{figure*}
\includegraphics [width=1. \linewidth]{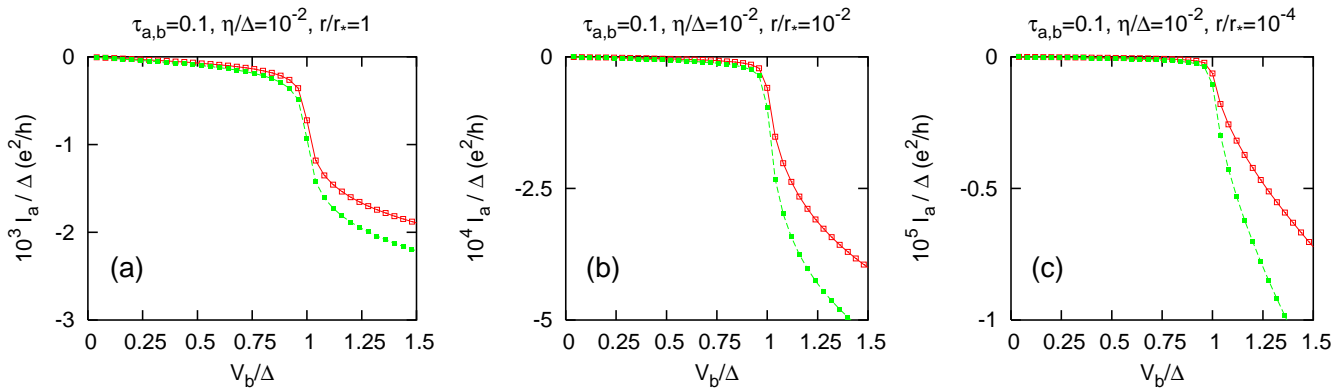}
\caption{(a) Variation of the crossed current $I_a(V_b)$
as a function of $V_b/\Delta$, for
$\eta/\Delta=10^{-2}$, $\tau_{a,b}=0.1$. The three panels
correspond to $r=1$ (a), $r=10^{-2}$ (b), and $r=10^{-4}$ (c).
The different curves correspond to $I_a(V_b)$
in the antiparallel alignment (red, $\Box$), and
to $I_a(V_b)$
in the parallel alignment (green, $\blacksquare$).
The absolute value of the crossed current is larger in
the parallel alignment.
}
\label{fig:I_Vqp}
\end{figure*}
\section{Numerical results}
\label{sec:num}
\subsection{FSF double junction with weak energy relaxation}
\label{sec:weak}
In the case of weak energy relaxation ($\tau_d \ll \tau_E \ll \tau_{sf}$)
we describe S' by a distribution function $f_{S'}(\omega)$,
supposed to be uniform in space within S', and determined
in such a way as to impose current conservation for each energy,
similarly to section~\ref{sec:normal-elas}.
Within the numerical approach we can treat the full voltage dependence
of the transmission coefficients, for arbitrary interface transparencies.
The quasiparticle transmission coefficients
deduced from Ref.~\onlinecite{Cuevas} are given in Appendix~\ref{app:qp}.

The variations of the sequential tunneling
crossed current $I_a(V_b)$ in the
parallel and antiparallel alignments are shown on Fig.~\ref{fig:elas-P}
for $\tau_{a,b}=0.1$ and $\tau_{a,b}=1$.
The absolute value of the
crossed current is larger in the parallel alignment
than in the antiparallel alignment (which coincides
with the normal state behavior).
For small values of $r/r_*$ the crossed conductance for
$e V_b > \Delta$ is almost equal to the crossed conductance
for $e V_b < \Delta$, in agreement with the argument
given in section~\ref{sec:normal-elas}.

\subsection{FSF double junction with strong energy relaxation}
\label{sec:strong}

The case of strong energy relaxation ($\tau_E\ll\tau_d\ll\tau_{sf}$)
can be obtained with low transparency interfaces in the trilayer geometry
since 
the transport dwell time can be sufficiently long in
this case. This case is likely to be irrelevant to the experiments by
Beckmann {\it et al.}\cite{Beckmann} involving highly transparent
FS interfaces, but is of interest for future experiments with
tunnel interfaces in the trilayer geometry.
The quasiparticle potentials $V_{qp}^{\uparrow}$ and
$V_{qp}^{\downarrow}$ are calculated self-consistently in
such a way that the integrated current
satisfies Kirchoff law separately in the
spin-up and spin-down channels.
For the sake of generality we do not treat only the case
$r/r_* \agt 1$, relevant to the trilayer geometry,
but use also $r/r_* \ll 1$.

The variation of the normalized
self-consistent quasiparticle potentials $e V_{qp}^\uparrow/\Delta$
and $e V_{qp}^\downarrow/\Delta$ as a function of $V/\Delta$ are
shown on Fig.~\ref{fig:Vqp} for different values of $r$ and
for small interface transparencies ($\tau_{a,b}=0.1$).
For $r/r_*=1$, the normalized quasiparticle potentials 
$e V_{qp}^\uparrow/\Delta$
and $e V_{qp}^\downarrow/\Delta$
increase from $0$ to a value close to unity as $e V_b$ increases
from $0$ to $\Delta$. The quasiparticle potentials are much reduced
as $r/r_*$ decreases, due to the fact that the superconductor S
tends to reduce the quasiparticle potentials in S' as the number
of channels of the SS' contact increases.
The quasiparticle potentials for $r/r_*=10^{-2}$ and
$r/r_*=10^{-4}$ are mostly determined by
the current flowing from S to S' and from S' to the ferromagnet ``b''.
We thus obtain $V_{qp}^{\uparrow,P}(V_b)/\Delta
\simeq V_{qp}^{\downarrow,AP}(V_b)/\Delta$
and $V_{qp}^{\uparrow,AP}(V_b)/\Delta
\simeq V_{qp}^{\downarrow,P}(V_b)/\Delta$, where P and
AP correspond to the parallel and antiparallel alignments.
These relations are due to the fact that the orientation of the
ferromagnet ``b'' is reversed when going from the parallel (P)
to the antiparallel (AP) spin orientation.

The variation of the crossed current $I_a(V_b)$ as a function
of $V_b$ is shown on Fig.~\ref{fig:I_Vqp} for the same parameters
as on Fig.~\ref{fig:Vqp}. The subgap crossed current becomes
very small for
small values of $r/r_*$, due to the fact that the self-consistent
quasiparticle potentials are also very small. The absolute value
of the crossed current is larger in the parallel
alignment than in the parallel alignment.

\section{Conclusion}
\label{sec:conclu}

To conclude we have discussed a model of
sequential tunneling crossed conductance
based on out-of-equilibrium spin populations in the
superconductor. 
The case of strong energy relaxation is 
expected to be realized for small interface transparencies 
in the trilayer geometry\cite{Takahashi,Zheng,Bozovic,Yamashita,Brataas},
where the
transport dwell time can be larger than the energy relaxation time.
In the other case of highly transparent interfaces corresponding to
the experiment by Beckmann {\it et al.}\cite{Beckmann},
the transport dwell
time is expected to be much smaller so that elastic transport can apply.
The geometrical effects are encoded in
a parameter $r$, very small in the geometry with lateral
contacts of the experiment by Beckmann {\it et
al.}\cite{Beckmann}. Another small parameter is $\eta/\Delta$,
proportional to the residual density of states within the
superconducting gap. There exists a cross-over value $r_*$
such that for $r \ll r_*$ the crossed conductance is almost
voltage-independent in the elastic model. The values of
$\eta/\Delta$ used in the literature\cite{Cuevas,Pekola}
indicate that the condition $r \ll r_*$ is verified in
the experiment by Beckmann {\it et al.}\cite{Beckmann}.
It is found  experimentally that for the
smallest distance between the ferromagnets the crossed
conductance is the same in the superconducting and normal
phases, which can be successfully reproduced by the
model of sequential tunneling.
Therefore a very small residual density of state
within the superconducting gap can lead to a sequential
tunneling current in the geometry with lateral contacts,
compatible with experiments, which is our main conclusion.
In the regime of strong energy relaxation, not relevant to
the experiments by Beckmann {\it et al.}\cite{Beckmann},
the crossed
conductance above the superconducting gap is much larger than the
crossed conductance below the superconducting gap. Contrary to
the elastic case, there exists a peak in the crossed conductance
for $e V_b \simeq \Delta$ (corresponding to a large slope in
the crossed current $I_a(V_b)$ on Fig.~\ref{fig:I_Vqp}).

In the case of tunnel interfaces,
the crossed conductance due to the spatially separated processes
is positive in the antiparallel alignment\cite{Falci,MF-PRB} while 
the sequential tunneling crossed
conductance is negative.
Tunnel interfaces would thus constitute an experimental test of
the possible effects. Replacing the ferromagnets by normal
metals\cite{Russo} with tunnel interfaces
in the geometry used by Beckmann {\it et al.}\cite{Beckmann}
would also be of interest.
On the theoretical side it would be useful to investigate
the spatial dependence of the out-of-equilibrium phenomena,
and investigate a more microscopic
model in which the size of the out-of-equilibrium
region would be controlled by the inverse proximity effect.
It would be also interesting
to use quasi-classical theory for describing a diffusive
superconductor\cite{Belzig}.

\section*{Acknowledgments}
The author thanks D. Feinberg for numerous discussions
on related problems, and thanks H. Courtois for
a critical reading of the manuscript. The author also
benefited from useful comments by F. Pistolesi, and a
from a fruitful discussion with B. Pannetier.

\appendix

\section{Transmission coefficients of a FS interface}
\label{app:FS}
\label{app:qp}
The different terms contributing to subgap and quasiparticle transport
at a FS interface are derived 
in Ref.~\onlinecite{Cuevas} by Keldysh Green's
function methods. In this Appendix we just recall these results
and assume non equilibrium distribution functions in the
superconductor.

A first term in the spin-up quasiparticle current
corresponds to transmission without branch crossing:
\begin{equation}
\label{eq:Iqp1}
I_{\rm qp,e\uparrow}^{(1)} = 
\int T_{e,\uparrow,{\rm loc}}^{(1)}(\omega)
\left[ f_{S'}^{(\uparrow)}(\omega)-n_F(\omega-e V_a)(\omega)\right]
d\omega
,
\end{equation}
with
\begin{equation}
\label{eq:Teup1}
T_{e,\uparrow,{\rm loc}}^{(1)}(\omega)=4 \pi^2 t_a^2 \frac{e}{h}
\rho_{a,a}^{1,1} \rho_{\alpha,\alpha}^{1,1}(\omega)
\left|1+t_a G_{a,\alpha}^{1,1,A}(\omega)\right|^2
.
\end{equation}

A second term in the quasiparticle current
corresponds to transmission with branch crossing\cite{BTK}:
\begin{equation}
I_{\rm qp,e\uparrow}^{(2)} =
\int T_{e,\uparrow,{\rm loc}}^{(2)}(\omega)
\left[ f_{S'}^{(\uparrow)}(\omega)-n_F(\omega+e V_a) \right]
d\omega
\label{eq:Iqp2}
,
\end{equation}
with
\begin{equation}
\label{eq:Teup2}
T_{e,\uparrow,{\rm loc}}^{(2)}(\omega) = 4 \pi^2 t_a^2 \frac{e}{h}
\rho_{a,a}^{2,2} \rho_{\alpha,\alpha}^{1,1}(\omega)
\left|t_a G_{a,\alpha}^{1,2,A}(\omega)\right|^2
.
\end{equation}
A spin-up electron from the ferromagnet ``a''
is transmitted in the superconductor while a Cooper pair
is annihilated in the superconductor therefore producing a net
transfer of a spin-down hole in the superconductor.

The third term in the quasiparticle current
is given by
\begin{equation}
\label{eq:Iqp3}
I_{\rm qp,e\uparrow}^{(3)} = \int
T_{e,\uparrow,{\rm loc}}^{(3)}(\omega)
\left[ n_F(\omega-e V_a)-n_F(\omega+e V_a) \right]
d\omega
,
\end{equation}
with
\begin{eqnarray}
\label{eq:Teup3}
T_{e,\uparrow,{\rm loc}}^{(3)}(\omega) &=&
-4 \pi^2 t_a^3 \frac{e}{h}
\rho_{a,a}^{1,1} \rho_{\alpha,\alpha}^{1,2}(\omega)\\
&\times& \mbox{Re} \left\{ \left[1+t_a G_{\alpha,a}^{1,1,R}(\omega)
\right] G_{a,\alpha}^{2,1,A}(\omega) \right\}
.
\nonumber
\end{eqnarray}
The density of state prefactors in
the first term of Eq.~(\ref{eq:Iqp3}) are given by
$\rho_{a,\uparrow} \rho_{a,\downarrow} 
\rho_{\alpha,\alpha}^{1,2}$ if
we use $G_{a,\alpha}^{2,1,A}=- i \pi t_a \rho_{a,a}^{2,2}
G_{\alpha,\alpha}^{2,1,A}$.
This process therefore corresponds to the
transmission of a spin-up electron from the ferromagnet to the
superconductor. At the same time a particle-hole excitation
is created at the
interface, the spin-down hole is backscattered in the ferromagnet
and the spin-up electron is transmitted is the superconductor.
This process in the quasiparticle channel
is reminiscent of the Andreev reflection term.

\end{document}